\documentclass[reprint,superscriptaddress,amsmath,amssymb,prb]{revtex4-2}
\usepackage{graphicx} 
\usepackage{dcolumn}  
\usepackage{bm}       
\usepackage{amsmath}
\usepackage{amssymb}
\usepackage{hyperref}
\usepackage{color,xcolor}
\usepackage{epstopdf}
\usepackage{ifpdf}

\begin{document}

\preprint{APS/123-QED}

\title{Hydrothermal synthesis and complete phase diagram of FeSe$_{1-x}$S$_{x}$ $(0 \leq x \leq 1)$ single crystals}

\author{Xiaolei Yi}
\affiliation{School of Physics, Southeast University, Nanjing 211189, China}
\author{Xiangzhuo Xing}
\email{xzxing@seu.edu.cn}
\affiliation{School of Physics, Southeast University, Nanjing 211189, China}
\author{Lingyao Qin}
\affiliation{School of Physics, Southeast University, Nanjing 211189, China}
\author{Jiajia Feng}
\affiliation{School of Physics, Southeast University, Nanjing 211189, China}
\author{Meng Li}
\affiliation{School of Physics, Southeast University, Nanjing 211189, China}
\author{Yufeng Zhang}
\affiliation{School of Physics, Southeast University, Nanjing 211189, China}
\author{Yan Meng}
\affiliation{School of Physics, Southeast University, Nanjing 211189, China}
\author{Nan Zhou}
\affiliation{School of Physics, Southeast University, Nanjing 211189, China}
\author{Yue Sun}
\email{sunyue@phys.aoyama.ac.jp}
\affiliation{School of Physics, Southeast University, Nanjing 211189, China}
\affiliation{Department of Physics and Mathematics, Aoyama Gakuin University, Sagamihara 252-5258, Japan}
\author{Zhixiang Shi}
\email{zxshi@seu.edu.cn}
\affiliation{School of Physics, Southeast University, Nanjing 211189, China}

\date{\today}

\begin{abstract}
We report the successful synthesis of FeSe$_{1-x}$S$_{x}$ single crystals with $x$ ranging from 0 to 1 via a hydrothermal method. A complete phase diagram of FeSe$_{1-x}$S$_{x}$ has been obtained based on resistivity and magnetization measurements. The nematicity is  suppressed with increasing $x$, and a small superconducting dome appears within the nematic phase. Outside the nematic phase, the superconductivity is continuously suppressed and reaches a minimum $T_\textup{c}$ at $x$ = 0.45; beyond this point, $T_\textup{c}$ slowly increases until $x$ = 1. Intriguingly, an anomalous resistivity upturn with a characteristic temperature $T^*$ in the intermediate region of $0.31 \leq x \leq 0.71$ is observed. $T^{*}$ shows a dome-like behavior with a maximum value at $x$ = 0.45, which is opposite the evolution of $T_\textup{c}$, indicating competition between $T^*$ and superconductivity. The origin of $T^*$ is discussed in detail. Furthermore, the normal state resistivity evolves from non-Fermi-liquid to Fermi-liquid behavior with S doping at low temperatures, accompanied by a reduction in electronic correlations. Our study addresses the lack of single crystals in the high-S doping region and provides a complete phase diagram, which will promote the study of relations among nematicity, superconductivity, and magnetism.
\end{abstract}

\maketitle

\section{Introduction}
The discovery of superconductivity in iron-based superconductors (IBSs) \cite{Hosono 1111} has opened a new era in the search for high-temperature superconductors and in research on the mechanisms governing unconventional superconductivity. The iron chalcogenide superconductor FeSe \cite{M. K. Wu FeSe}, which exhibits the simplest crystal structure among IBSs, has attracted considerable interest due to its unique properties \cite{Shibauchi JPSJ review}. FeSe is a compensated semimetal that exhibits superconductivity with a superconducting (SC) transition temperature $T_\textup{c}\sim 9$\, K at ambient pressure \cite{M. K. Wu FeSe}. Intriguingly, $T_\textup{c}$ can be increased to 37 K by the application of pressure \cite{HP 37K}, to values exceeding 40 K through intercalation \cite{X. F. Lu}, ionic liquid gating \cite{B. Lei}, and potassium deposition \cite{k dosing}. More surprisingly, signs of superconductivity with $T_\textup{c}$ exceeding 65 K have also been observed in a monolayer FeSe film on a SrTiO$_3$ substrate \cite{FeSe 65 K,FeSe 100 K}. Because the Fermi energies are extremely small and comparable to the SC gap, the superconductivity in bulk FeSe has been argued to be close to a Bardeen-Cooper-Schrieffer/Bose-Einstein-condensation crossover \cite{S. Kasahara-10}. Unlike other IBSs, FeSe undergoes a structural (nematic) transition at $T_\textup{s} \sim 87$\, K without a magnetic transition \cite{Shibauchi JPSJ review,T. M. McQueen-11,S. H. Baek-12,A. E. Bohmer-13,Coldea REVIEW-14}. Therefore, FeSe is a fascinating platform for investigating the interplay between nematicity and superconductivity.

Isovalent S substitution, which maintains the nature of compensated semimetals, is an effective route for tuning the ground state and electronic interactions in FeSe. It has been reported that the nematic order is gradually suppressed by S doping in FeSe$_{1-x}$S$_{x}$ \cite{Coldea REVIEW-14,ncp-15,P. Wiecki-16,Coldea-17,Watson-18}, and a nonmagnetic nematic quantum critical point (QCP) appears at $x \sim 0.17$ \cite{ncp-15}. Nuclear magnetic resonance measurements have revealed that antiferromagnetic (AFM) fluctuations are slightly enhanced and then strongly suppressed with S substitution, leading to negligible AFM fluctuations near the QCP \cite{P. Wiecki-16}. The nematic QCP has a significant impact on both the SC \cite{Sato-19,Hanaguri-20} and normal state properties \cite{Bristow-21,Huang-22,nature-23}. The SC gap exhibits an abrupt change across the nematic QCP \cite{Sato-19}. Within the nematic phase, the small gap is highly anisotropic with deep minima or nodes, while the large gap is more isotropic \cite{Sato-19,Sun-24,Sun-25,Feng D L-26,Sprau-27}. By contrast, the larger gap becomes strongly anisotropic outside the nematic phase \cite{Sato-19,Hanaguri-20}. Meanwhile, the normal state resistivity in the vicinity of the QCP exhibits $T$-linear behavior at low temperatures \cite{Bristow-21,Huang-22,nature-23,Lederera-28}, signifying non-Fermi-liquid behavior due to the nematic critical fluctuation. Moreover, quantum oscillation measurements have indicated a topological Lifshitz transition and a reduction in electronic correlations across the nematic QCP \cite{Coldea-17}.

Thus far, although increasingly more peculiarities of FeSe$_{1-x}$S$_{x}$ have been revealed, very few studies have reported on the high-S doping region due to difficulties in synthesizing FeSe$_{1-x}$S$_{x}$ single crystals with $x > 0.29$ by chemical vapor transport (CVT). Recently, Nabeshima \emph{et al.} reported the growth of FeSe$_{1-x}$S$_{x}$ thin films up to $x = 0.43$ \cite{film jpsj-29}. However, these thin films show some differences from CVT-grown single crystals, possibly due to the lattice strain effect. To get deeper insights into the evolution of intrinsic bulk properties, the synthesis of FeSe$_{1-x}$S$_{x}$ single crystals with a higher S content ($x > 0.29$) is highly desirable.

In this paper, we report the successful synthesis of a series of FeSe$_{1-x}$S$_{x}$ single crystals with S content covering the full range $(0 \leq x \leq 1)$. The single crystals were produced by a hydrothermal method, which enabled us to obtain a complete doping phase diagram. We find that the phase diagram in the low-S doping region agrees with that established from CVT-grown single crystals. Outside the nematic phase, the superconductivity is continuously suppressed and reaches a minimum $T_\textup{c} = 2.8$\, K at $x = 0.45$; beyond this point, $T_c$ slowly increases toward that of FeS. Remarkably, an anomalous resistivity upturn above $T_\textup{c}$ is observed with a characteristic temperature $T^*$, which exhibits dome-shaped dependence on S content and a maximum value at $x = 0.45$, indicating competition between $T_\textup{c}$ and $T^*$. Furthermore, the normal state resistivity evolves from non-Fermi-liquid to Fermi-liquid behavior with S doping at low temperatures, accompanied by a reduction in electronic correlations.

\section{Experimental details}
FeSe$_{1-x}$S$_{x}$ single crystals were synthesized by deintercalation of K ions from K$_{0.8}$Fe$_{1.6}$Se$_{2-x}$S$_x$ precursors using a hydrothermal ion release/introduction technique \cite{Dong x l-30,cpb yuan-31,linhai-32,FeS-HC}, as depicted by a schematic diagram in Fig. 1(a). First, K$_{0.8}$Fe$_{1.6}$Se$_{2-x}$S$_x$ single crystals were synthesized by a self-flux method as described in the Supplemental Material \cite{supplemental-32}. For the hydrothermal reactions, a given amount of NaOH was dissolved in 10\, mL of deionized water in a Teflon-lined stainless-steel autoclave (25\, mL). Then, Fe powder, selenourea, thiourea, and several pieces of K$_{0.8}$Fe$_{1.6}$Se$_{2-x}$S$_x$ single crystals were added to the solution. Details regarding the quantities of starting materials for the different FeSe$_{1-x}$S$_{x}$ single crystals are given in Table S1 \cite{supplemental-32}. The autoclave was tightly sealed, heated to $403 - 423$ \, K, and maintained for $50 - 70$ h. FeSe$_{1-x}$S$_{x}$ single crystals were finally obtained by washing the products with deionized water and were found to be almost identical to the K$_{0.8}$Fe$_{1.6}$Se$_{2-x}$S$_x$ precursors in shape and size, as shown in Fig. 1(b).

Single-crystal x-ray diffraction (XRD) measurements were performed at room temperature on a commercial Rigaku diffractometer with Cu $K\alpha$ radiation, and elemental analysis was performed by a scanning electron microscope equipped with an energy dispersive x-ray (EDX) spectroscopy probe. Electrical transport and specific heat measurements were performed on a physical property measurement system (PPMS-9T, Quantum Design). Magnetization measurements were also performed on a physical property measurement system (PPMS-9T, Quantum Design) with a vibrating sample magnetometer attachment.

\section{Results and Discussion}

\begin{figure*}
\centering
\includegraphics[width=0.95\textwidth]{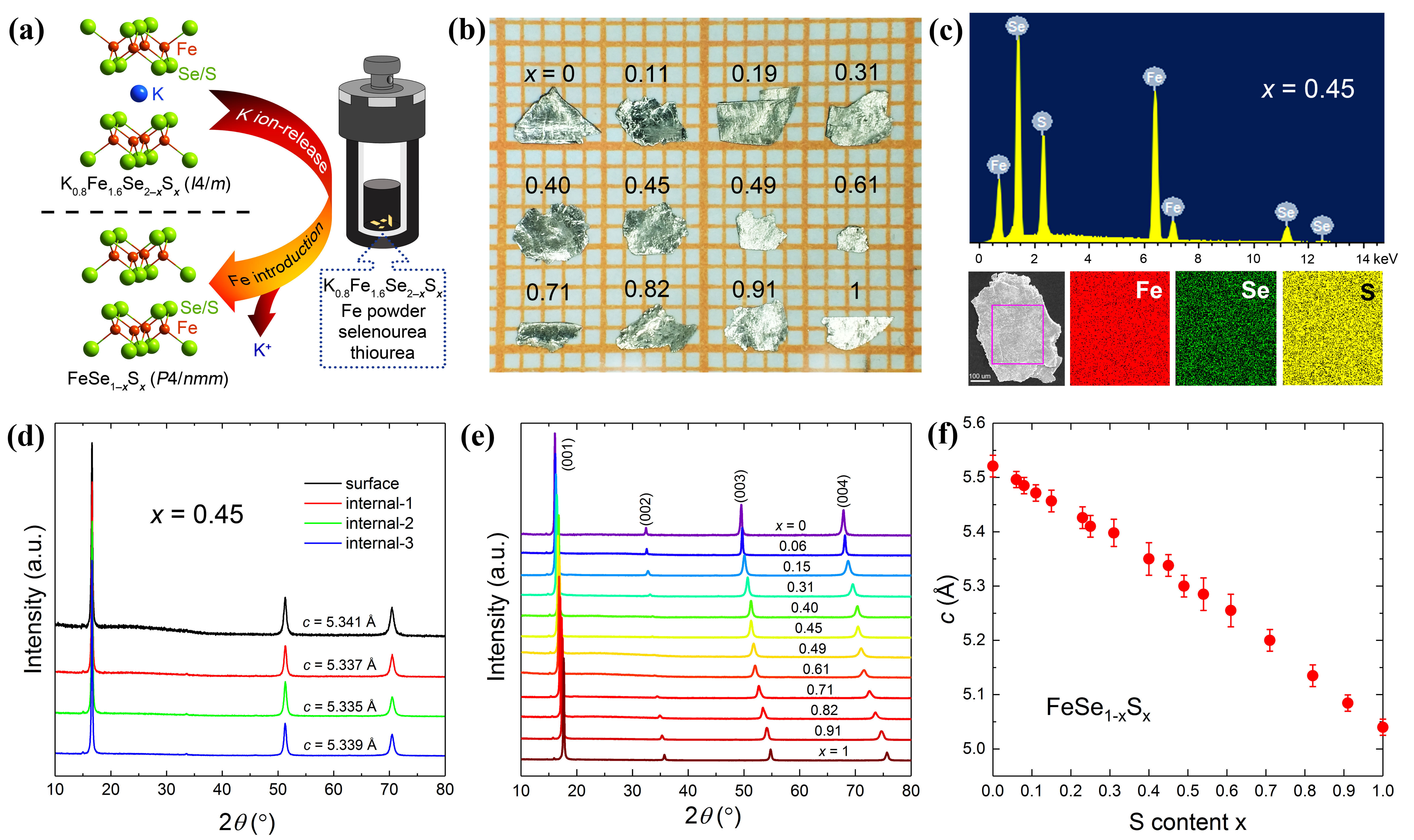}
\caption{\label{fig1}(a) Schematic illustration of the hydrothermal ion release/introduction route for the synthesis of FeSe$_{1-x}$S$_{x}$ single crystals. (b) Optical image of select FeSe$_{1-x}$S$_{x}$ single crystals. (c) EDX spectrum (top panel) and compositional mappings (bottom panels) of Fe, Se, and S in the selected rectangular region for a representative FeSe$_{0.55}$S$_{0.45}$ single crystal. (d) Single-crystal XRD patterns measured on different parts of the FeSe$_{0.55}$S$_{0.45}$ single crystal. The one marked ``surface'' is measured on the external surface, and the others, marked ``internal,'' are measured on three inside surfaces prepared by cleaving three times on the same crystal. (e) Single-crystal XRD patterns of FeSe$_{1-x}$S$_{x}$ single crystals with various values of $x$ after the cleavage. All (00\emph{l}) peaks obviously shift to higher angles with increasing $x$, indicative of effective S doping. (f) Lattice parameter $c$ deduced from the XRD patterns as a function of S content $x$. }
\end{figure*}
The actual S content of the obtained FeSe$_{1-x}$S$_{x}$ single crystals was determined by EDX measurements. Several representative EDX spectra are presented in the top panel of Fig. 1(c) and  Fig. S1 \cite{supplemental-32}. The results show that no trace of K is detected for any of the single crystals, indicating that the interlayer K ions are completely released from the K$_{0.8}$Fe$_{1.6}$Se$_{2-x}$S$_x$ precursors. The bottom panel of Fig. 1(c) shows the compositional mappings of the FeSe$_{0.55}$S$_{0.45}$ single crystal, performed on the inner surface after the cleavage, signifying that Fe, Se, and S are almost homogeneously distributed in the crystal. For each single crystal, about $10 - 15$ different spots were selected in the EDX measurements, and the average was used to determine the actual stoichiometry composition, as listed in Table S1 \cite{supplemental-32}. It is found that the actual S content is very close to that of K$_{0.8}$Fe$_{1.6}$Se$_{2-x}$S$_x$ precursors, indicating that the S dopants occupying the Se sites remain almost unchanged after the hydrothermal process. Henceforth, we will refer to the single crystals by their actual S content throughout this paper.

We performed the XRD measurements on different parts of the obtained FeSe$_{1-x}$S$_{x}$ single crystals. The typical results of a representative crystal with $x = 0.45$ are shown in Fig. 1(d). The XRD patterns marked ``surface'' and ``internal'' denote the ones taken from the external surface of the crystal directly and the inner surfaces prepared by continuously cleaving three times in the same crystal, respectively. Only the (00\emph{l}) peaks are detected, indicating the good $c$-axis orientation of the obtained single crystals. Meanwhile, all the peaks can be well indexed by a tetragonal structure with space group $P4/nmm$, in contrast to the case of K$_{0.8}$Fe$_{1.6}$Se$_{2-x}$S$_x$ with $I4/m$ shown in Fig. S2(a) \cite{supplemental-32}. In addition, the peaks for the surface and inner parts are almost identical, yielding the same $c$-axis lattice constant of 5.338(3)\,{\AA}, which proves the uniform property of our single crystals. Figure 1(e) shows the single-crystal XRD patterns of FeSe$_{1-x}$S$_{x}$ single crystals with different values of $x$ after the cleavage. By refining the (00\emph{l}) peaks, the lattice parameter $c$ as a function of $x$ is obtained, as shown in Fig. 1(f). As expected, the lattice parameter $c$ decreases monotonically with increasing $x$ due to the smaller ionic radius of S$^{2-}$ compared with Se$^{2-}$, indicating that the S atoms are successfully incorporated into the crystal lattice.

\begin{figure*}
\begin{center}
\includegraphics[width=1\textwidth]{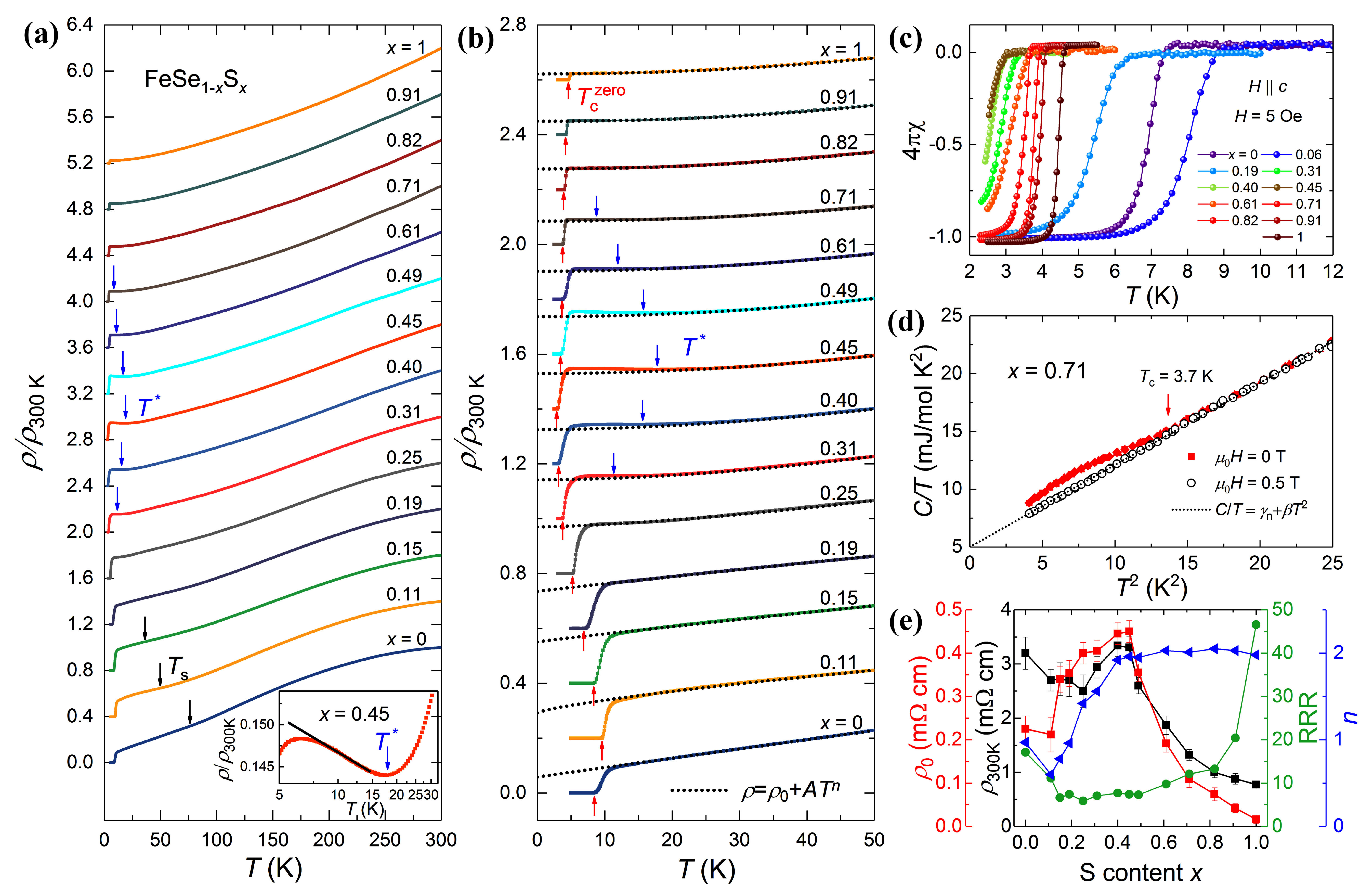}
\end{center}
\caption{\label{fig2}(a) Temperature dependence of the in-plane resistivity (normalized by corresponding values at $T = 300$ K) for FeSe$_{1-x}$S$_{x}$ ($0 \leq x \leq 1$) single crystals. The inset shows the normalized resistivity of a representative crystal with $x = 0.45$ near the upturn region in log scale. The straight line is a guide to the eyes. (b) Enlarged view of the normalized resistivity for temperatures below $T = 50$\, K. The data are vertically shifted for clarity. The arrows indicate the nematic transition at $T_\textup{s}$, resistivity minimum at $T^*$, and SC transition at $T^\textup{zero}_c$. The black dotted lines in (b) represent fits to the formula $\rho(T)=\rho_{0}+AT^{n}$ (see text for details). (c) Temperature dependence of ZFC magnetization for FeSe$_{1-x}$S$_{x}$ single crystals under 5\, Oe with $H \parallel c$. (d) Low-temperature specific heat measured at $H$ = 0 T and 0.5 T for the crystal with $x$ = 0.71. The arrow indicates the onset of a superconducting transition. (e) Room-temperature resistivity $\rho_\textup{{300\,K}}$, residual resistivity $\rho_0$, temperature exponent $n$, and the residual resistivity ratio (RRR) as a function of S content, $x$. Error bars represent statistical standard deviation for three to five crystals of each composition.}
\end{figure*}

Figure 2(a) presents the temperature dependence of the in-plane resistivity $\rho(T)$ normalized to the corresponding value at 300\, K [see Fig. 2(e)] for different values of $x$. For undoped FeSe single crystals, the nematic transition occurs near $T_\textup{s} \sim 75$\, K, as determined from the temperature derivative of resistivity ($d\rho$/$dT$) shown in Fig. S2(b) \cite{supplemental-32}, which is somewhat lower than that of CVT-grown samples \cite{ncp-15,CVT-34,Bohmer-35,Sun-36}. With S doping, $T_s$ is gradually suppressed. For $x < 0.31$, FeSe$_{1-x}$S$_{x}$ single crystals exhibit metallic behavior over the entire temperature range. Extending to higher doping levels, a resistive anomaly with a characteristic temperature, $T^*$, emerges and survives up to $x \sim 0.71$. Below $T^*$, the resistivity shows an anomalous upturn and obeys a logarithmic temperature dependence before the SC transition, as is clearly shown in the inset of Fig. 2(a) for a representative example with $x = 0.45$. Data for the other crystals can be seen in Fig. S2(c) \cite{supplemental-32}. $T^*$ shows a nonmonotonic doping dependence; that is, $T^*$ initially increases and then decreases with S doping. It is noted that a similar resistivity upturn has also been recently observed in FeSe$_{1-x}$S$_{x}$ ($x \leq 0.43$) thin films \cite{film jpsj-29}, possibly due to a magnetic transition, as revealed in FeSe and FeSe$_{1-x}$S$_{x}$ single crystals under  pressure \cite{Sun J P-37,Matsuura-38,Terashima-39}. Details of the origin of this upturn behavior will be discussed later.

\begin{figure*}
\begin{center}
\includegraphics[width=1\textwidth]{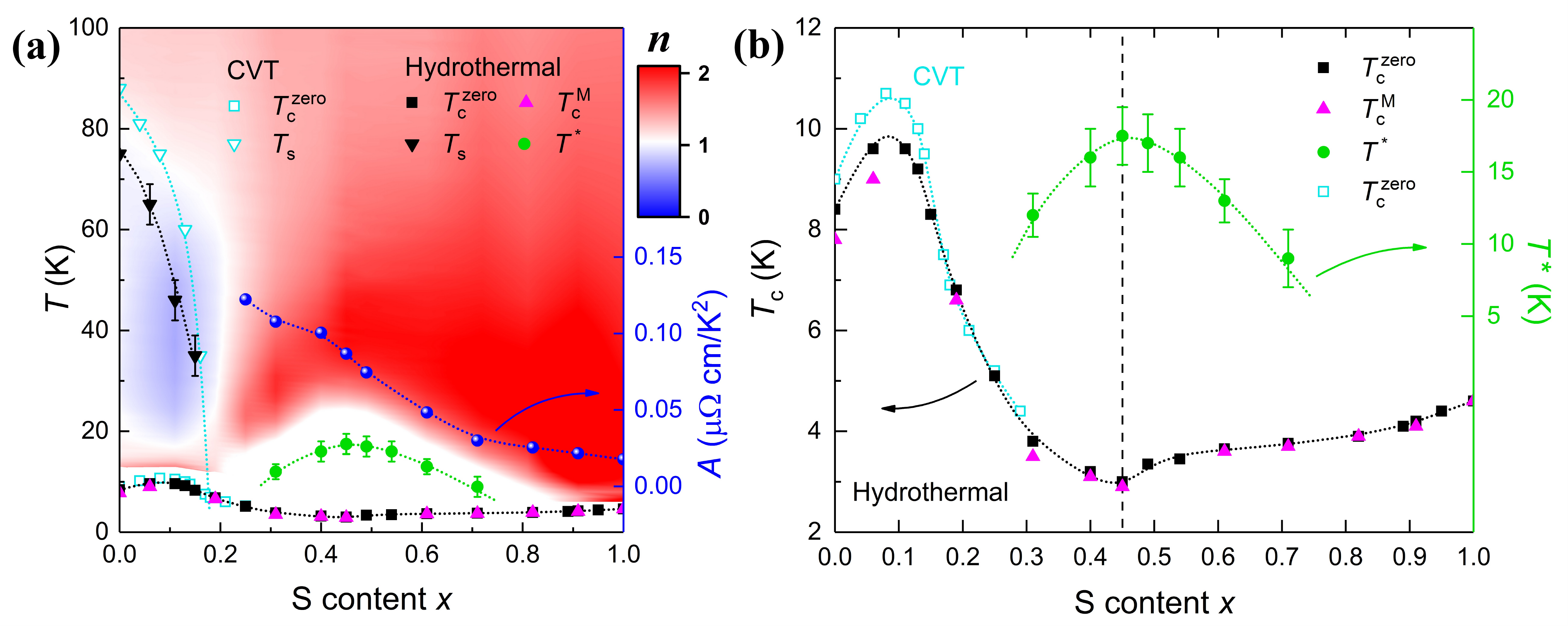}
\end{center}
\caption{\label{fig3}(a) Complete phase diagram of hydrothermal FeSe$_{1-x}$S$_{x}$ single crystals in the present study. $T_\textup{s}$ represents the nematic transition temperature. $T^\textup{zero}_\textup{c}$ and $T^\textup{M}_\textup{c}$ are the SC transition temperatures obtained from resistivity and magnetization measurements, respectively. $T^*$ is the characteristic temperature at which the $\rho-T$ curves show local minima at low temperatures. For comparison, corresponding data for CVT-grown FeSe$_{1-x}$S$_{x}$ ($0 \leq x \leq 0.29$) single crystals reported in Refs. \cite{ncp-15,P. Wiecki-16,Bristow-21} are also included. The temperature dependence of the exponent $n$ extracted from $d\textup{ln}(\rho-\rho_0)/d\textup{ln}T$ for each crystal is represented as a contour map (see text). The coefficient $A$ is obtained by fitting the low-temperature resistivity using the Fermi-liquid formula $\rho(T)=\rho_{0}+AT^{2}$ (see Fig. S3 \cite{supplemental-32}). (b) Enlarged view of $T_\textup{c}$ and $T^*$ as a function of S content $x$.}
\end{figure*}

Figure 2(b) shows an enlarged view of normalized $\rho(T)$ curves below $T = 50$ K. The SC transition temperature, $T^\textup{zero}_\textup{c}$, defined as the zero-resistivity temperature, is indicated by red arrows. With increasing $x$, $T^\textup{zero}_\textup{c}$ increases slightly from 8.4\, K in FeSe to 9.6\, K at $x \sim 0.11$, followed by a decrease to a minimum value of 2.8\, K at $x = 0.45$. Then, $T^\textup{zero}_\textup{c}$ gradually increases to 4.6\, K in FeS ($x = 1$). Figure 2(c) shows the temperature dependence of the zero-field-cooled (ZFC) magnetization under an applied magnetic field of 5\, Oe with $H \parallel c$. $T^\textup{M}_\textup{c}$ is defined as the onset of diamagnetism, which is almost consistent with the $T^\textup{zero}_\textup{c}$ value determined from the resistivity measurements. $4\pi \chi$ was calculated after considering the demagnetization effect, which saturates at low temperatures and reaches around -1 (indicating the superconducting volume fraction is $\sim$ 100$\%$) for $x \leq 0.19$ and $x \geq 0.71$. In the intermediate region, $4\pi \chi$ locates in the range of -0.35 to -0.85 at 2\, K, which may saturate and reach -1 at $T < 2$ K. Except for $x = 0.45$ with the lowest $T_c$, the significant volume fractions estimated at 2 K indicate the bulk superconductivity. To further confirm whether the superconductivity is bulk or not, we performed the specific heat measurement on a representative crystal with $x$ = 0.71. Figure 2(d) shows the temperature dependence of the specific heat, plotted as $C/T$ versus $T^2$. At zero field, a clear specific heat jump associated with the superconducting transition appears at $T_c$ = 3.7\, K, consistent with the magnetic susceptibility and resistivity measurements, which confirms the nature of bulk superconductivity. A weak magnetic field 0.5 T is enough to suppress the superconducting jump, similar to that reported in the FeS single crystal synthesized by the same hydrothermal technique \cite{FeS-HC,FeS Xingjie}.

To gain more insight into the effect of doping on the normal state transport properties, particularly in the S doping region with $x > 0.25$, we analyzed the normal state resistivity below $T = 50$\, K by fitting the results to the power law $\rho(T)=\rho_{0}+AT^{n}$, where $\rho_0$ is the residual resistivity at zero temperature and $n$ is the temperature exponent \cite{Analytis-40,Luo H q-41,Kasahara-42}. For $0.31 \leq x \leq 0.71$, the fits are performed in the temperature region of $T^*+10$\, K to 50\, K to minimize the effect of upturn behavior. The fitting curves are represented by black dotted lines in Fig. 2(b). The obtained exponent $n$ and residual resistivity $\rho_0$ are plotted in Fig. 2(e). In addition, the residual resistivity ratio (RRR) value, which is defined as $\rho(\text{300 K})$/$\rho_0$ and describes the strength of disorder scattering in samples, is also shown in Fig. 2(e). The RRR firstly decreases from 17 in undoped FeSe single crystals to 6 at $x = 0.25$ (smaller than that of CVT-grown single crystals \cite{Sato-19,Bohmer-35,Sun-36}). The ratio slowly increases with further S doping and finally increases steeply to 48 for $x = 1$.

Typically, $n = 2$ is expected for a conventional metal described by the Landau Fermi-liquid theory, while a power law with $n < 2$ indicates non-Fermi-liquid behavior. As shown in Fig. 2(e), the exponent $n$ maintains a nearly constant value slightly below 1 in the low-S doping region. The exponent then increases until saturating at 2 with increasing $x$, indicating an evolution from non-Fermi-liquid to Fermi-liquid transport. This evolution can be more clearly seen in a contour plot of the temperature-dependent $n$ extracted from $d\textup{ln}(\rho-\rho_0)/d\textup{ln}T$ in Fig. 3(a). The resistivity shows a sublinear temperature dependence in the nematic phase. Outside the nematic phase, a dominant $T^{1.5}$ dependence is visible at high temperatures, and a Fermi-liquid behavior is recovered with decreasing temperature at low temperatures. This $T^{1.5}$ dependence was also observed in the CVT-grown single crystals with $x \leq 0.25$ \cite{Bristow-21}. As $x$ increases further, the Fermi-liquid region observed at low temperatures extends to higher temperatures.

In the Fermi-liquid region with $\rho(T)=\rho_{0}+AT^{2}$, the coefficient $A$ is proportional to the carrier's effective mass, i.e., $A \propto (m^*/m_0)^2$. Thus, we replotted the $\rho(T)$ data for $x \geq 0.25$ in the form of $\rho$ versus $T^2$, as shown in Fig. S3 \cite{supplemental-32}. The coefficient $A$ obtained from a linear fitting to low temperatures is presented in Fig. 3(a). Clearly, $A$ decreases monotonically with increasing $x$, signifying a reduction in effective mass with S doping. Indeed, angle-resolved photoemission spectroscopy indicates that the electronic correlations are suppressed by S doping in FeSe$_{1-x}$S$_{x}$ single crystals toward FeS, as evidenced by an increased Fermi velocity and bandwidth \cite{Reiss-43}.

Based on the resistivity and magnetization data, we constructed a complete phase diagram of FeSe$_{1-x}$S$_{x}$ single crystals, as shown in Fig. 3(a). For comparison, corresponding data obtained from the CVT-grown FeSe$_{1-x}$S$_{x}$ ($0 \leq x \leq 0.29$) single crystals reported in Refs. \cite{ncp-15,P. Wiecki-16,Bristow-21} are also included. With isovalent S doping, the nematic transition is found to be gradually suppressed and finally vanishes at $x \sim 0.17$. However, the superconductivity is robust against S doping and spans the entire doping region of $0 \leq x \leq 1$. To more clearly present the evolution of $T_\textup{c}$ with S doping, we redrew the phase diagram at low temperatures, as shown in Fig. 3(b). Within the nematic phase, a small dome-shaped SC phase is observed, consistent with previous results for CVT-grown single crystals \cite{Coldea REVIEW-14,ncp-15,P. Wiecki-16,Coldea-17}. It is noted that the values of both $T_\textup{s}$ and $T_\textup{c}$ in our hydrothermal single crystals are slightly lower than those of CVT-grown single crystals. B\"{o}hmer \emph{et al}. have argued that the reduction of both $T_\textup{s}$ and $T_\textup{c}$ correlates with decreasing RRR values, i.e., the disorder effect \cite{Bohmer-35}.

Outside the nematic phase, the superconductivity is continuously suppressed, leading to a minimum $T_\textup{c} = 2.8$\, K at $x = 0.45$; beyond this point, $T_\textup{c}$ slowly increases to 4.6\, K at $x = 1$. It has been reported that the SC gap exhibits an abrupt change across the nematic QCP ($x \sim 0.17$), suggesting that the pairing mechanism may be different inside (denoted as SC1) and outside (denoted as SC2) the nematic phase \cite{Sato-19,Hanaguri-20}. For $x > 0.45$, the trend of $T_\textup{c}$ suddenly reverses, and $T_\textup{c}$ slowly increases toward FeS, indicating that the SC phase at $x > 0.45$ may be different from that of SC2. A similar minimum $T_\textup{c}$ has also been observed in the middle of the phase diagram for other IBSs. In 1111-type LaFeAsO, a second SC dome has been reported at high doping levels upon substitution of O with F or H \cite{Iimura-47,CPL-48}. For 122-type KFe$_2$As$_2$, $T_\textup{c}$ first decreases and then increases with increasing pressure, exhibiting a V-shaped behavior, which may indicate a pairing symmetry change \cite{K122-49,Taufour-50}. Another remarkable feature of this phase diagram is the evolution of $T^*$ with S doping, which exhibits dome-like behavior with a maximum value of 17.5\, K at $x = 0.45$. $T^*$ has also been observed in FeSe$_{1-x}$S$_{x}$ ($x \leq 0.43$) thin films, emerging just after the disappearance of the nematic phase at $x \sim 0.18$ and increasing monotonically to $x = 0.43$ \cite{film jpsj-29}. In our single crystals, $T^*$ initially appears at $x = 0.31$, and its magnitude is smaller than that of thin films \cite{film jpsj-29}. Such differences may be due to the strain effect in thin films. Noticeably, the evolution of $T^*$ is opposite to that of $T_\textup{c}$, reflecting competition between $T_\textup{c}$ and $T^*$.

Next, we turn our attention to the origin of $T^*$. Considering the relatively small RRR values in the intermediate region of FeSe$_{1-x}$S$_{x}$ single crystals, the disorder effect may play an important role in the resistivity upturn. In the crystal with $x$ $\sim$ 0.5, the disorder effect would be the strongest due to the largest lattice mismatch in the Se/S layer [local structure distortion of Fe(Se,S)$_4$ tetrahedra]. Indeed, the room-temperature resistivity $\rho_\textup{{300\,K}}$ and residual resistivity $\rho_0$ manifest the maximum values at $x$ $\sim$ 0.5. In the disordered system, it is natural to deem that the resistivity upturn with logarithmic temperature dependence might be ascribed to the weak localization effect originating from the presence of disorder potentials \cite{Rev. Mod. Phys.-44}. If this is the case, the observation of negative magnetoresistance below $T^*$ is expected according to the theory of Anderson's localization \cite{Anserson theory,Anserson theory-2}. However, the resistivity increases upon applying magnetic field, as shown in Fig. S4 \cite{supplemental-32}, suggesting that the weak localization effect is unlikely to be applicable in our case.

Another possible origin of $T^*$ is related to the local magnetic impurity scattering, arising from defects at Fe sites. For example, in a Kondo system with conduction electrons and localized electrons caused by magnetic impurity, the Kondo singlet forms as the ground state. It will cause a Kondo behavior which also exhibits a resistivity upturn with logarithmic temperature dependence at low temperatures, as previously proposed in BaFe$_2$As$_2$ \cite{Ba122} and Ba(Fe$_{1-x}$Mn$_x$)$_2$As$_2$ \cite{BaMn-122}. Moreover, the upturn behavior was also observed in Fe$_{1+y}$Se$_{0.6}$Te$_{0.4}$ single crystals \cite{Sun-FeSeTe,Sun-sust}, which is attributed to the local magnetic moment caused by the excess Fe, leading to the localization of charge carriers. Another possible origin of  $T^*$ is the inelastic scattering due to spin fluctuations caused by crystallographic disorder. The resistivity upturn at low temperatures has been suggested to be related to spin fluctuations in BaFe$_2$As$_2$ system \cite{Ba122}. The spin fluctuations can be enhanced by magnetic field; that is, the resistivity minimum shifts to higher temperatures with increasing fields, consistent with our observation shown in Fig. S4 \cite{supplemental-32}. Certainly, further theoretical and experimental investigations are desired to clarify this issue in the FeSe$_{1-x}$S$_{x}$ system.

Finally, we discuss the relationship between $T^*$ and a possible magnetic transition which is reminiscent of the behavior of FeSe and FeSe$_{1-x}$S$_{x}$ single crystals under pressure \cite{Sun J P-37,Matsuura-38}, as proposed by Nabeshima \emph{et al}. for FeSe$_{1-x}$S$_{x}$ ($x \leq 0.43$) thin films \cite{film jpsj-29}. It has been demonstrated that the increase in chalcogen height $h_{Ch}$ may be responsible for the emergence of magnetic order in pressurized FeSe$_{1-x}$S$_{x}$ single crystals \cite{Moon-52}. However, in contrast to the physical pressure results, $h_{Ch}$ decreases monotonically upon isovalent S doping. As a consequence, the magnetic order shifts to higher pressures with increasing S content, as higher pressures are required for obtaining a larger $h_{Ch}$ to induce magnetic order  \cite{Matsuura-38}. In this sense, it may be difficult to stabilize the magnetic order by S doping at ambient pressure. Moreover, aside from the magnetism observed at high pressures \cite{Sun J P-37,Matsuura-38}, an additional low-pressure magnetic dome has also been detected in FeSe$_{1-x}$S$_{x}$ single crystals \cite{Xiang-53,Holenstein-54}. This magnetic order competes with superconductivity at low pressures, which leads to a local maximum in $T_\textup{c}$, where the magnetic order arises. It has been found that the local maximum in $T_\textup{c}$ shifts to lower pressures with increasing S doping, indicating the possible coexistence of superconductivity and magnetism in FeSe$_{1-x}$S$_{x}$ at ambient pressure for $x \geq 0.2$ \cite{Holenstein-54}. In order to check the above arguments, we have measured the temperature dependence of magnetic susceptibility under a field of 1 T applied in the $ab$ plane for several crystals with upturn behavior, as shown in Fig. S5 \cite{supplemental-32}. Clearly, all susceptibility curves show paramagnetic-like behavior, and no anomaly associated with the long-range magnetic order is observed at $T^*$. Consequently, the magnetic transition as observed in pressurized FeSe$_{1-x}$S$_{x}$ single crystals is unlikely to be responsible for the resistivity upturn at low temperatures.

\section{Conclusion}

In conclusion, we have synthesized a series of FeSe$_{1-x}$S$_{x}$ ($0 \leq x \leq 1$) single crystals by a hydrothermal method and have constructed a complete phase diagram based on resistivity and magnetization data. In the low-S doping region, the phase diagram is consistent with that of CVT-grown single crystals. Outside the nematic phase, the superconductivity is continuously suppressed and reaches a minimum $T_\textup{c}$ at $x = 0.45$; beyond this point, $T_\textup{c}$ slowly increases to $x = 1$. Remarkably, a resistivity upturn above $T_\textup{c}$ associated with a characteristic temperature $T^*$ is observed, which exhibits a dome shape with a maximum value at $x = 0.45$, indicative of competition between $T_\textup{c}$ and $T^*$. Local magnetic impurity scattering (e.g., the Kondo effect) or inelastic scattering (e.g., spin fluctuations) may be responsible for the resistivity anomaly at $T^*$.

\begin{acknowledgments}

This work was partly supported by the National Key R$\&$D Program of China (Grant No.\ 2018YFA0704300), the Strategic Priority Research Program (B) of the Chinese Academy of Sciences (Grant No.\ XDB25000000) and the National Natural Science Foundation of China (Grants No.\ U1932217 and No. 11674054). X. X. was also supported by a project funded by the China Postdoctoral Science Foundation (Grant No.\ 2019M661679) and the Jiangsu Planned Projects for Postdoctoral Research Funds (Grant No.\ 2019K149). Y. S. was supported by JSPS KAKENHI (Grant Nos.\ JP20H05164 and JP19K14661).

X. Y. and X. X. contributed equally to this work.
\end{acknowledgments}

\appendix


\begin{thebibliography}{199}
\bibitem{Hosono 1111}Y. Kamihara, T. Watanabe, M. Hirano, and H. Hosono, J. Am. Chem. Soc. \textbf {130}, 3296 (2008).
\bibitem{M. K. Wu FeSe}F. C. Hsu, J. Y. Luo, K. W. Yeh, T. K. Chen, T. W. Huang, P. M. Wu, Y. C. Lee, Y. L. Huang, Y. Y. Chu, D. C. Yan, and M. K. Wu, Proc. Natl. Acad. Sci. USA \textbf{105}, 14262 (2008).
\bibitem{Shibauchi JPSJ review}T. Shibauchi, T. Hanaguri, and Y. Matsuda, J. Phys. Soc. Jpn. \textbf{89}, 102002 (2020).
\bibitem{HP 37K}S. Medvedev, T. M. McQueen, I. A. Troyan, T. Palasyuk, M. I. Eremets, R. J. Cava, S. Naghavi, F. Casper, V. Ksenofontov, G.Wortmann, and C. Felser, Nat. Mater. \textbf{8}, 630 (2009).
\bibitem{X. F. Lu}X. F. Lu, N. Z. Wang, H. Wu, Y. P. Wu, D. Zhao, X. Z. Zeng, X. G. Luo, T. Wu, W. Bao, G. H. Zhang, F. Q. Huang, Q. Z. Huang, and X. H. Chen, Nat. Mater. \textbf{14}, 325-329 (2015).
\bibitem{B. Lei}B. Lei, J. H. Cui, Z. J. Xiang, C. Shang, N. Z. Wang, G. J. Ye, X. G. Luo, T. Wu, Z. Sun, and X. H. Chen, Phys. Rev. Lett. \textbf{116}, 077002 (2016).
\bibitem{k dosing}C. H. Wen, H. C. Xu, C. Chen, Z. C. Huang, X. Lou, Y. J. Pu, Q. Song, B. P. Xie, M. Abdel-Hafiez, D. A. Chareev, A. N. Vasiliev, R. Peng, and D. L. Feng, Nat. Commun. \textbf{7}, 10840 (2016).
\bibitem{FeSe 100 K}J. F. Ge, Z. L. Liu, C. Liu, C. L. Gao, D. Qian, Q. K. Xue, Y. Liu, and J. F. Jia, Nat. Mater. \textbf{14}, 285-289 (2015).
\bibitem{FeSe 65 K}Q.-Y. Wang, Z. Li, W.-H. Zhang, Z.-C. Zhang, J.-S. Zhang, W. Li, H. Ding, Y.-B. Ou, P. Deng, K. Chang, J. Wen, C.-L. Song, K. He, J.-F. Jia, S.-H. Ji, Y.-Y. Wang, L.-L. Wang, X. Chen, X.-C. Ma, and Q.-K. Xue, Chin. Phys. Lett. \textbf{29}, 037402 (2012).
\bibitem{S. Kasahara-10}S. Kasahara, T. Watashige, T. Hanaguri, Y. Kohsaka, T. Yamashita, Y. Shimoyama, Y. Mizukami, R. Endo, H. Ikeda, K. Aoyama, T. Terashima, S. Uji, T. Wolf, H. von Lohneysen, T. Shibauchi, and Y. Matsuda, Proc. Natl. Acad. Sci. USA \textbf{111}, 16309 (2014).
\bibitem{T. M. McQueen-11}T. M. McQueen, A. J. Williams, P. W. Stephens, J. Tao, Y. Zhu, V. Ksenofontov, F. Casper, C. Felser, and R. J. Cava, Phys. Rev. Lett. \textbf{103}, 057002 (2009).
\bibitem{S. H. Baek-12}S. H. Baek, D. V. Efremov, J. M. Ok, J. S. Kim, J. van den Brink, and B. Buchner, Nat. Mater. \textbf{14}, 210 (2015).
\bibitem{A. E. Bohmer-13}A. E. Bohmer, T. Arai, F. Hardy, T. Hattori, T. Iye, T. Wolf, H. V. Lohneysen, K. Ishida, and C. Meingast, Phys. Rev. Lett. \textbf{114}, 027001 (2015).
\bibitem{Coldea REVIEW-14}A. I. Coldea, Front. Phys. \textbf{8}, 594500 (2021).
\bibitem{ncp-15}S. Hosoi, K. Matsuura, K. Ishida, H. Wang, Y. Mizukami, T. Watashige, S. Kasahara, Y. Matsuda, and T. Shibauchi, Proc. Natl. Acad. Sci. USA \textbf{113}, 8139-8143 (2016).
\bibitem{P. Wiecki-16}P. Wiecki, K. Rana, A. E. B\"{o}hmer, Y. Lee, S. L. Bud'ko, P. C. Canfield, and Y. Furukawa, phys. Rev. B \textbf{98}, 020507(R) (2018).
\bibitem{Coldea-17}A. I. Coldea, S. F. Blake, S. Kasahara, A. A. Haghighirad, M. D. Watson, W. Knafo, E. S. Choi, A. McCollam, P. Reiss, T. Yamashita, M. Bruma, S. C. Speller, Y. Matsuda, T. Wolf, T. Shibauchi, and A. J. Schofield, npj Quan. Mater. \textbf{4}, 2 (2019).
\bibitem{Watson-18}M. D. Watson, T. K. Kim, A. A. Haghighirad, S. F. Blake, N. R. Davies, M. Hoesch, T. Wolf, and A. I. Coldea, Phys. Rev. B \textbf{92}, 121108(R) (2015).
\bibitem{Sato-19}Y. Sato, S. Kasahara, T. Taniguchi, X. Xing, Y. Kasahara, Y. Tokiwa, Y. Yamakawa, H. Kontani, T. Shibauchi, and Y. Matsuda, Proc. Natl. Acad. Sci. USA \textbf{115}, 1227 (2018).
\bibitem{Hanaguri-20}T. Hanaguri, Y. Kohsaka, T. Machida, T. Watashige, S. Kasahara, T. Shibauchi, and Y. Matsuda, Sci. Adv. \textbf{4}, eaar6419 (2018).
\bibitem{Bristow-21}M. Bristow, P. Reiss, A. A. Haghighirad, Z. Zajicek, S. J. Singh, T. Wolf, D. Graf, W. Knafo, A. McCollam, and A. I. Coldea, Phys. Rev. Research \textbf{2}, 013309 (2020).
\bibitem{Huang-22}W. K. Huang, S. Hosoi, M. \v{C}ulo, S. Kasahara, Y. Sato, K. Matsuura, Y. Mizukami, M. Berben, N. E. Hussey, H. Kontani, T. Shibauchi, and Y. Matsuda, Phys. Rev. Research \textbf{2}, 033367 (2020).
\bibitem{nature-23}S. Licciardello, J. Buhot, J. Lu, J. Ayres, S. Kasahara, Y. Matsuda, T. Shibauchi, and N. E. Hussey, Nature \textbf{567}, 213-217 (2019).
\bibitem{Sun-24}Y. Sun, A. Park, S. Pyon, T. Tamegai, and H. Kitamura, Phys. Rev. B \textbf{96}, 140505(R) (2017).
\bibitem{Sun-25}Y. Sun, S. Kittaka, S. Nakamura, T. Sakakibara, K. Irie, T. Nomoto, K. Machida, J. Chen, and T. Tamegai, Phys. Rev. B \textbf{96}, 220505(R) (2017).
\bibitem{Feng D L-26}H. C. Xu, X. H. Niu, D. F. Xu, J. Jiang, Q. Yao, Q. Y. Chen, Q. Song, M. Abdel-Hafiez, D. A. Chareev, A. N. Vasiliev, Q. S. Wang, H. L. Wo, J. Zhao, R. Peng, and D. L. Feng, Phys. Rev. Lett. \textbf{117}, 157003 (2016).
\bibitem{Sprau-27}P. O. Sprau, A. Kostin, A. Kreisel, A. E. B\"{o}hmer, V. Taufour, P. C. Canfield, S. Mukherjee, P. J. Hirschfeld, B. M. Andersen, and J. C. S. Davis, Science \textbf{357}, 75-80 (2017).
\bibitem{Lederera-28}S. Lederera, Y. Schattnerb, E. Berg, and S. A. Kivelsond, Proc. Natl Acad. Sci. USA \textbf{114}, 4905 (2017).
\bibitem{film jpsj-29}F. Nabeshima, T. Ishikawa, K.-i. Oyanagi, M. Kawai, and A. Maeda, J. Phys. Soc. Jpn. \textbf{87}, 073704 (2018).
\bibitem{Dong x l-30}X. Dong, K. Jin, D. Yuan, H. Zhou, J. Yuan, Y. Huang, W. Hua, J. Sun, P. Zheng, W. Hu, Y. Mao, M. Ma, G. Zhang, F. Zhou, and Z. Zhao, Phys. Rev. B \textbf{92}, 064515 (2015).
\bibitem{cpb yuan-31}D. Yuan, Y. Huang, S. Ni, H. Zhou, Y. Mao, W. Hu, J. Yuan, K. Jin, G. Zhang, X. Dong, and F. Zhou, Chin. Phys. B \textbf{25}, 077404 (2016).
\bibitem{linhai-32}H. Lin, Y. Li, Q. Deng, J. Xing, J. Liu, X. Zhu, H. Yang, and H. H. Wen, Phys. Rev. B \textbf{93}, 144505 (2016).
\bibitem{FeS-HC}C. K. H. Borg, X. Zhou, C. Eckberg, D. J. Campbell, S. R. Saha, J. Paglione, and E. E. Rodriguez, Phys. Rev. B \textbf{93}, 094522(2016).
\bibitem{supplemental-32}See Supplemental Material for more details about the crystal growth and characterization of K$_{0.8}$Fe$_{1.6}$Se$_{2-x}$S$_x$ precursors, compositional analysis, determination of $T_s$ and quadratic temperature coefficient $A$, resistivity upturn and its field response, and the normal state magnetic susceptibility.
\bibitem{CVT-34}D. Chareev, Y. Ovchenkov, L. Shvanskaya, A. Kovalskii, M. Abdel-Hafiez, D. J. Trainer, E. M. Lechner, M. Iavarone, O. Volkova, and A. Vasiliev, CrystEngComm \textbf{20}, 2449 (2018).
\bibitem{Bohmer-35}A. E. B\"{o}hmer, V. Taufour, W. E. Straszheim, T. Wolf, and P. C. Canfield, Phys. Rev. B \textbf{94}, 024526 (2016).
\bibitem{Sun-36}Y. Sun, S. Pyon, and T. Tamegai, Phys. Rev. B \textbf{93}, 104502 (2016).
\bibitem{Sun J P-37}J. P. Sun, K. Matsuura, G. Z. Ye, Y. Mizukami, M. Shimozawa, K. Matsubayashi, M. Yamashita, T. Watashige, S. Kasahara, Y. Matsuda, J. Q. Yan, B. C. Sales, Y. Uwatoko, J. G. Cheng, and T. Shibauchi, Nat. Commun. \textbf{7}, 12146 (2016).
\bibitem{Matsuura-38}K. Matsuura, Y. Mizukami, Y. Arai, Y. Sugimura, N. Maejima, A. Machida, T. Watanuki, T. Fukuda, T. Yajima, Z. Hiroi, K. Y. Yip, Y. C. Chan, Q. Niu, S. Hosoi, K. Ishida, K. Mukasa, S. Kasahara, J. G. Cheng, S. K. Goh, Y. Matsuda, Y. Uwatoko, and T. Shibauchi, Nat. Commun. \textbf{8}, 1143 (2017).
\bibitem{Terashima-39}T. Terashima, N. Kikugawa, S. Kasahara, T. Watashige, T. Shibauchi, Y. Matsuda, T. Wolf, A. E. B\"{o}hmer, F. Hardy, C. Meingast, H. v. L\"{o}hneysen, and S. Uji, J. Phys. Soc. Jpn. \textbf{84}, 063701 (2015).
\bibitem{FeS Xingjie}J. Xing, H. Lin, Y. Li, S. Li, X. Zhu, H. Yang, and H. H. Wen, Phys. Rev. B \textbf{93}, 104520 (2016).
\bibitem{Analytis-40}J. G. Analytis, H.-H. Kuo, R. D. McDonald, MarkWartenbe, P. M. C. Rourke, N. E. Hussey, and I. R. Fisher, Nat. Phys. \textbf{10}, 194 (2014).
\bibitem{Luo H q-41}R. Zhang, D. Gong, X. Lu, S. Li, M. Laver, C. Niedermayer, S. Danilkin, G. Deng, P. Dai, and H. Luo, Phys. Rev. B \textbf{91}, 094506 (2015).
\bibitem{Kasahara-42}S. Kasahara, T. Shibauchi, K. Hashimoto, K. Ikada, S. Tonegawa, R. Okazaki, H. Shishido, H. Ikeda, H. Takeya, K. Hirata, T. Terashima, and Y. Matsuda, Phys. Rev. B \textbf{81}, 184519 (2010).
\bibitem{Reiss-43}P. Reiss, M. D. Watson, T. K. Kim, A. A. Haghighirad, D. N. Woodruff, M. Bruma, S. J. Clarke, and A. I. Coldea, Phys. Rev. B \textbf{96}, 121103 (2017).
\bibitem{Iimura-47}S. Iimura, S. Matsuishi, H. Sato, T. Hanna, Y. Muraba, S. W. Kim, J. E. Kim, M. Takata, and H. Hosono, Nat. Commun. \textbf{3}, 943 (2012).
\bibitem{CPL-48}J. Yang, R. Zhou, L.-L. Wei, H.-X. Yang, J.-Q. Li, Z.-X. Zhao, and G.-Q. Zheng, Chin. Phys. Lett. \textbf{32}, 107401 (2015).
\bibitem{K122-49}F. F. Tafti, A. Juneau-Fecteau, M.-\`{E}. Delage, S. R. d. Cotret, J.-P. Reid, A. F. Wang, X.-G. Luo, X. H. Chen, N. Doiron-Leyraud, and L. Taillefer, Nat. Phys. \textbf{9}, 349 (2013).
\bibitem{Taufour-50}V. Taufour, N. Foroozani, M. A. Tanatar, J. Lim, U. Kaluarachchi, S. K. Kim, Y. Liu, T. A. Lograsso, V. G. Kogan, R. Prozorov, S. L. Bud'ko, J. S. Schilling, and P. C. Canfield, Phys. Rev. B \textbf{89}, 220509(R) (2014).
\bibitem{Rev. Mod. Phys.-44}P. A. Lee, and T. V. Ramakrishnan, Rev. Mod. Phys. \textbf{57}, 287-337 (1985).
\bibitem{Anserson theory} P. W. Anderson, Phys. Rev. \textbf{109}, 1492 (1958).
\bibitem{Anserson theory-2} P. A. Lee, and D. S. Fisher, Phys. Rev. Lett. \textbf{47}, 882 (1981).
\bibitem{Ba122} X. F. Wang, T. Wu, G. Wu, H. Chen, Y. L. Xie, J. J. Ying, Y. J. Yan, R. H. Liu, and X. H. Chen, Phys. Rev. Lett. \textbf{102}, 117005 (2009).
\bibitem{BaMn-122}T. Urata, Y. Tanabe, K. K. Huynh, H. Oguro, K. Watanabe, S. Heguri, and K. Tanigaki, Phys. Rev. B \textbf{89}, 024503 (2014).
\bibitem{Sun-FeSeTe}Y. Sun, T. Taen, T. Yamada, S. Pyon, T. Nishizaki, Z. Shi, and T. Tamegai, Phys. Rev. B \textbf{89}, 144512 (2014).
\bibitem{Sun-sust}Y. Sun, Z. Shi, and T. Tamegai, Supercond. Sci. Technol. \textbf{32}, 103001 (2019).
\bibitem{Moon-52}C. Y. Moon, and H. J. Choi, Phys. Rev. Lett. \textbf{104}, 057003 (2010).
\bibitem{Xiang-53}L. Xiang, U. S. Kaluarachchi, A. E. B\"{o}hmer, V. Taufour, M. A. Tanatar, R. Prozorov, S. L. Bud'ko, and P. C. Canfield, Phys. Rev. B \textbf{96}, 024511 (2017).
\bibitem{Holenstein-54} S. Holenstein, J. Stahl, Z. Shermadini, G. Simutis, V. Grinenko, D. A. Chareev, R. Khasanov, J. C. Orain, A. Amato, H. H. Klauss, E. Morenzoni, D. Johrendt, and H. Luetkens, Phys. Rev. Lett. \textbf{123}, 147001 (2019).




\end{thebibliography}
\end{document}